**Introducing a Cost-Effective Approach for Improving the Arterial Traffic Performance Operating Under the Semi-Actuated Coordinated Signal Control**

**Sina Dabiri, (Corresponding Author)**
Ph.D. Student
Charles Edward Via, Jr. Department of Civil and Environmental Engineering,
Virginia polytechnic institute and state university
750 Drillfield Drive, 301 Patton Hall, Blacksburg, VA 24061
Tel: (540) 231-6635; Fax: (540) 231-7532; Email: sina@vt.edu

**Kianoush Kompany**
Traffic Engineer
Atkins North America
1616 East Millbrook Road, Suite 160, Raleigh, NC 27609
Tel: (540) 998-5015; Email: kianoush.kompany@atkinsglobal.com

**Montasir Abbas**
Associate Professor
Charles Edward Via, Jr. Department of Civil and Environmental Engineering,
Virginia polytechnic institute and state university
750 Drillfield Drive, 301-A Patton Hall, Blacksburg, VA 24061
Tel: (540) 231-9002; Fax: (540) 231-7532; Email: abbas@vt.edu

Word count:  5,385 words text + 7 tables/figures x 250 words (each) = 7,135 words

Submission Date: 11/24/2017

**ABSTRACT**

The semi-actuated coordinated operation mode is a type of signal control where minor approaches are placed with detectors to develop actuated phasing while major movements are coordinated without using detection systems. The objective of this study is to propose a cost-effective approach for reducing delay in the semi-actuated coordinated signal operation without incurring any extra costs in terms of installing new detectors or developing adaptive controller systems. We propose a simple approach for further enhancing a pre-optimized timing plan. In this method, the green splits of non-coordinated phases are multiplied by a factor greater than one. In the meantime, the amount of green time added to the non-coordinated phases is subtracted from the coordinated phases to keep the cycle length constant. Thus, if the traffic demand on the side streets exceeds the expected traffic flow, the added time in the non-coordinated phase enables the non-coordinated phases to accommodate the additional traffic demand. A regression analysis is implemented so as to identify the optimal value of the mentioned factor, called Actuated Factor (ActF). The response variable is the average delay reduction (seconds/vehicle) of the simulation runs under the proposed signal timing plan compared to the simulation runs under the pre-optimized timing plan, obtained through the macroscopic signal optimization tools. External traffic movements, left-turn percentage, and ActF are the explanatory variables in the model. Results reveal that the ActF is the only significant variable with the optimal value of 1.15 that is applicable for a wide range of traffic volumes.

# INTRODUCTION

Implementing optimal signal timing plans at intersections is one of the effective solutions for managing traffic operations. By convention, a proper method of traffic signal control is selected depending on traffic characteristics and roadway geometry conditions. Traffic signal control schemes usually involve a pre-timed or an actuated mode, or a combination of the two (*1*). As opposed to the pre-timed mode, the actuated control dynamically responds to the fluctuations in the traffic demand information collected by detectors. The actuated operation is divided into two categories: semi-actuated and fully actuated. In the semi-actuated signal control, the detectors are installed only on the minor traffic movements. Green time is assigned by default to the major approaches unless a call is placed on one of the minor approaches. Coordination is another strategy for programming the traffic signals, which provides a progressive traffic flow between multiple closely spaced intersections along an arterial to reduce the traffic performance indices such as the number of stops, delay, and travel time (*1*) Establishing a temporal relationship between the green times of similar phases in the adjacent arterial intersections so that the vehicles move in platoons with a minimum number of stops is the main purpose of a coordinated system. Integrating semi-actuated and coordinated operation modes generates a semi-actuated coordinated control architecture which outperforms the semi-actuated and fully actuated signal timing systems. Such a method engenders significant improvements of vehicular travel times and stops (*2, 3*).

In arterials that the speed limits on the coordinated streets are less than 40 mph, coordinated phases on the major street can turn green without using any detector, while non-coordinated green phases, on the side streets, are requested through actuations of detection systems. The green splits of coordinated phases are at least corresponding to the split time in the optimal timing plan. However, the signal states for non-coordinated phases remain green unless their designated detectors gap out or max out. The unused green time due to either gap-out or phase skip of the non-coordinated phases is transferred to the coordinated phases, which returns the green to the coordinated phases earlier than the normal setup. Sophisticated understanding of the factors and parameters related to the semi-actuated coordinated signal operation offers us an opportunity to improve such a traffic signal control with a simple yet efficient method.

To date, several studies have investigated various semi-actuated coordinated parameters in order to suggest the optimal method for implementing this type of control on arterial traffic networks. In addition to the major signal timing parameters, (i.e., cycle length, green splits, offsets, and phase sequence), an actuated phase contains more attributes such as minimum green time, vehicle extension, and maximum green time. A lot of macroscopic and primarily deterministic programs such as VISTRO, Synchro, and TRANSYT-7F are available to optimize major signal timing settings for arterial intersections. Microscopic simulation optimization techniques have also been introduced to not only address the incapability of macroscopic models in considering the stochastic nature of traffic and the drivers' interactions but also optimize the actuated control parameters (*4*). Nonetheless, overwhelming computation time has still remained as a shortcoming in the simulation-based optimization models. Detection configuration is another necessary part of the actuated controls to place a call when a vehicle is detected, extend the green time, and identify gaps between vehicle actuations. Several published studies have reported the best detection elements including the detectors' operation mode, memory mode, location, and length of the detectors in accordance with the designated traffic signal control scheme and the mean speed of traffic movements (*1, 5*). A key issue regarding the detection design is that changing the design configuration incurs an extra cost by adding or replacing the detectors with brand new ones, which the authorities may not agree to implement. The above-mentioned limitations on the arterial semi-actuated coordinated traffic controls call for developing a methodology that improves the measures of effectiveness in the network without incurring additional costs, in a reasonable computation time.

To address the above-mentioned limitations, we introduce a simple and straightforward procedure for enhancing the optimal signal timing plan for arterials that practitioners have already decided to be operated with the semi-actuated operation mode. The optimal signal timing plan have been already obtained with macroscopic traffic analysis tools. In this study, the actuated parameters and detection elements are configured according to available traffic guidelines and manuals, whereas the major optimal signal timing parameters are generated using VISTRO, as a powerful macroscopic signal optimization software package. While offsets, phase sequence, and background cycle length are kept constant, the green splits for non-coordinated phases are multiplied by a factor greater than one. Note that no detectors are utilized for the coordinated movements since the operation mode is semi-actuated. The extra green time of non-coordinated phases and the multiplication factor are called Added Green Time (AGT) and Actuated Factor (ActF), respectively. Then, to preserve the background cycle length, AGT is subtracted from the green time of the coordinated phases. The specific objective of this study is to identify the applicability of using such a factor (ActF) and evaluate its optimal value in different traffic conditions. The proposed methodology creates a signal timing plan so that the total delay in an arterial is reduced compared to the optimal timing plan produced by VISTRO. Furthermore, we assess the significance of the major and side streets traffic volumes in determining the optimal ActF.

We call our methodology simple and cost-effective due to two principal reasons: 1) It does not incur additional costs for installing new detectors, changing detection configurations, or developing new software to dynamically optimize signal timing plans. 2) The recommended method requires only basic math calculation, which can be easily adopted by practitioners. The approach even does not need to deploy data-driven and advanced learning techniques that have recently been utilized a lot in transportation fields (6-8).

## ACTUATED AND COORDINATED CONTROL LITERATURE REVIEW

So far, a considerable amount of research has been conducted on a wide range of actuated and coordinated control aspects since the study of Jovanis et al., who introduced this type of traffic signal control *(9)*. A majority of methodologies for achieving an optimal actuated and coordinated operation can be categorized in two groups. The methods in the first group concentrated on the arterial control strategies while the second group focused on optimizing the signal timing parameters. Each of these viewpoints is discussed in the following paragraphs.

A summary of recommended guidelines and strategies for implementing the actuated-coordinated signal systems has been provided by Chang et al. *(10)*. In addition to the detector configuration and intersection timing parameters, the study provided a step-by-step procedure for selecting the optimal strategy including pre-timed conversion, early return, and v/c ratio control strategy. The users are able to select the appropriate strategy based on the level of saturation and the importance of the previous timing plan settings. Developing alternatives to improve an arterial performance using the actuated and coordinated concepts was the main objective of the research conducted by Messer et al. *(11)*. First, the optimal pre-timed coordinated timing plan was transformed into the actuated-coordinated timing plan via vehicle actuations. The effectiveness of this change was evaluated by traffic simulation. The results indicated that reducing the green time of the optimal pre-timed coordinated phases result in no performance improvement of the actuated system, whereas increasing the green split of the pre-timed coordinated phase deteriorated the performance. The rationale behind it is that the extra green time of the coordinated phases is not transferred to non-coordinated phases when demand in the minor street is high since detection systems are not installed on the major street to implement the gap-out operation. To propose an optimal strategy for one-way or two-way progression in the congested or uncongested direction, Chaudhary et al. assessed five coordination strategies by applying a simulation study when the degree of saturation for each intersection varies from 0.86 to 1.12 *(12)*. Findings revealed that the one-way progression in the congested direction was an appropriate method when the network was experiencing oversaturation. Abbas et al. outlined a framework for the design of arterial control strategies for oversaturated conditions *(13)*. The set of network critical links and their traffic volumes were essential factors in determining the appropriate control strategy. In a study by He et al., multiple priority requests from various modes of transportation (e.g., emergency vehicles, buses, and pedestrians) were accommodated in a coordinated and actuated signal control system (14). Considering an isolated semi-actuated intersection, Lin et al. investigated the effect of side-street traffic volumes on the major-street green time ratio. They sought to find a relationship between traffic volumes and number of stops as the base for making the optimal decision in signal coordination (15). In comparison with adaptive control strategies, Chia et al. demonstrated that conventional actuated-coordinated time-of-day plans work better during the peak hours (16).

Assessing the actuated and coordinated control parameters was the focus of the second category. Early return to green and the formation of the stop bar waiting queues occur frequently in the actuated-coordinated signal systems. Consequently, designing a constant offset on the basis of free-flow speed and the arterial link length causes a considerable amount of delay, which reveals the need for an adaptive offset algorithm. To overcome this shortcoming, a real-time offset transitioning algorithm was proposed to dynamically adjust the offset with the purpose of providing smooth progression of the vehicles moving on the corridor *(17)*. The issue of early return to green can also be addressed by considering the stochastic distribution of starts/ends of green for the coordinated phases according to a large signal state data obtained from a field data collection operation *(18)*. Taking the advantage of these distributions, an offline offset refiner was proposed to fine-tune the offset after designing the signal timing plans. Defining force-off points is a common method for running the coordination in the actuated-coordinated signal operation. Comparing the performance of the floating and fixed force-off modes for various volume-to-capacity ratios revealed that the fixed force-off mode outperformed the other modes in all levels of the traffic demand *(19)*. In the fixed force-off mode, in contrast to the floating force-off mode, the available unused green time due to gapping out or phase skipping of the non-coordinated phases can be added to the next phase, given that the traffic volume on the next phase is excessive. No matter whether the next phase is coordinated or not, "it may use the extra time up to the force-off point" (*1*). However, in the floating force-off mode, the only upcoming movement that can utilize the unused green time, of the extra green time that gapped out earlier, is the coordinated phase. Therefore, floating force-off limits the flexibility of the amount of the green times assigned to the non-coordinated phases (*1*). In addition to benefits of traditional actuated-coordinated signal systems, a neuro-fuzzy signal controller can also improve the progression by establishing a real-time secondary coordination between the upstream coordinated through-movement, and the downstream non-coordinated left-turn movement *(20)*. The proposed controller allows for alleviating queue spillback and upstream link blockage at closely-spaced intersections, which are the two traffic issues that stem from the early return to green and the demand variation of the side-street links.

Inability of the commercial macroscopic traffic signal optimization tools for taking into account the traffic uncertainties urges the need for developing microscopic simulation-based traffic signal optimization methods *(21)*. Moreover, these existing optimization software packages primarily focus on major timing plan parameters, (e.g., green splits) rather than actuated parameters, (e.g., vehicle extension). Considering the retrospective approximation concept, Markov monotonic search algorithm as the optimizer, and VISSIM as the simulation engine, a framework for optimizing the maximum green time as one of the actuated parameters was developed *(22)*. Park et al. presented a stochastic-optimization method for simultaneously optimizing the basic signal timing parameters and actuated settings including green time, vehicle extension, and recall mode *(4, 23)*. The Genetic Algorithm (GA) and Shuffled Frog-Leaping Algorithm were used as the optimizers, which were interfaced with CORSIM as the microscopic simulation platform.

Although many studies have been conducted to investigate various aspects of the actuated-coordinated systems, no simple and practical methodology has been proposed to improve the available optimal signal timing plan in a practically feasible computation time. To address such a lack of research, we seek to propose a straightforward methodology on the basis of the signal timing plan so as to ameliorate the arterial performance.

**METHODOLOGY**
In the semi-actuated coordinated systems, detectors are only installed along the non-coordinated movements on the side streets and the left-turn movements on the major street. Non-coordinated phases serve for at least their designated minimum green time only if the controller receives a call from their associated detectors; otherwise, the phase is skipped. The green time for the actuated phases is extended by a defined unit extension value if the gap between two consecutive vehicle actuations is less than the vehicle extension parameter; if not, the phase is gapped out. The maximum green time, which is equal to the optimal green split created by an optimization tool, is reached if there is continuous demand on the actuated phase. The unused green time of the actuated phases due to gapping out or phase skipping is added to the green time of coordinated phases. Thus, the green split of the coordinated phases may be extended by the unused green time of the actuated phases. To synchronize the coordinated phases, the controller is forced to operate within the constraint of a constant cycle length.

Using the concept of the semi-actuated coordinated signal control, which means increasing the green split in the non-coordinated phases while keeping the cycle length constant, might be an efficient solution for improving the arterial performance. In case of demand fluctuations on the side streets, increasing the green split time of the side streets also gives room to handle the extra traffic demand of the side streets more effectively. In case that there is no extra demand, the additional green time would be transferred to the coordinated phases. It should be noted that the increase in the green splits of the non-coordinated phases requires to be accompanied by the decrease in the green splits of the coordinated phases so as to maintain the designated cycle length constant. Short-term traffic fluctuations within an hour need to be considered using Peak Hour Factor (PHF), which is the ratio of the average hourly volume to four times the peak 15- minute volume. In the absence of the field measurements, a value of 0.92 is recommended for the urban area by the Highway Capacity Manual, which is employed in this study *(24)*.

To implement the proposed method, first of all, the optimal timing plan parameters including offsets, cycle length, and green splits are determined using VISTRO, a commercial macroscopic simulation-optimization tool with a high computational speed. In this study, pedestrian phases are not considered during the development of the timing plans. In the next step, the green splits of the actuated non-coordinated phases, calculated by VISTRO for each intersection, are multiplied by a factor greater than one, which is called Actuated Factor (ActF). The optimal value of ActF is determined by a regression analysis, which is explained in the following sections. Then, the total added seconds in each ring diagram are deducted from the coordinated phase in that ring in order to keep the cycle length constant. Similar to the cycle length, the offsets of the new timing plan are the same as the VISTRO timing plan. For the sake of simplicity, the green splits of the timing plan created by VISTRO and the new timing plan are called "Base TP" and "Factored TP", respectively. Employing the NEMA phase numbering scheme and the standard ring-barrier diagram, the Factored TP is determined as follows:

$$g_{i,\text{Factored TP}} = ActF * g_{i,\text{Base TP}} \quad \forall i \in \text{non - coordinated phases} \quad (1)$$

$$g_{j1,\text{Factored TP}} = g_{j1,\text{Base TP}} - \left(\sum_i g_{i,\text{Factored TP}} - \sum_i g_{i,\text{Base TP}}\right) \quad i \in \text{non - coordinated phases in ring 1} \quad (2)$$

$$g_{j2,\text{Factored TP}} = g_{j2,\text{Base TP}} - \left(\sum_i g_{i,\text{Factored TP}} - \sum_i g_{i,\text{Base TP}}\right) \quad i \in \text{non - coordinated phases in ring 2,} \quad (3)$$

Where $g_i$ is the green time of the non-coordinated phase $i$ in the corresponding timing plan and ActF is the actuated factor that the non-coordinated phases need to be multiplied by. $g_{j1}$ and $g_{j2}$ denote the green splits of coordinated phases in ring 1 and 2, respectively. Equation (1) shows how to convert the green splits of the non-coordinated phases in the Base TP to the Factored TP while equations (2) and (3) represent how to obtain the new values for the coordinated phases in rings 1 and 2, respectively. It is worth mentioning that if the amount of $g_{j1,\text{Factored TP}}$ and $g_{j2,\text{Factored TP}}$ become lower than their associated minimum green, an adjustment must be applied to keep all green splits equal to or greater than their minimum values. The procedure of obtaining Factored TP is illustrated through an example in Figure 1 by using equations (1) to (3) when the ActF is 1.10. The first and the second diagrams show the Base TP and Factored TP, respectively, with a cycle length of 60 seconds. In this example, phases 2 (in ring 1) and 6 (in ring 2) are the coordinated phases.

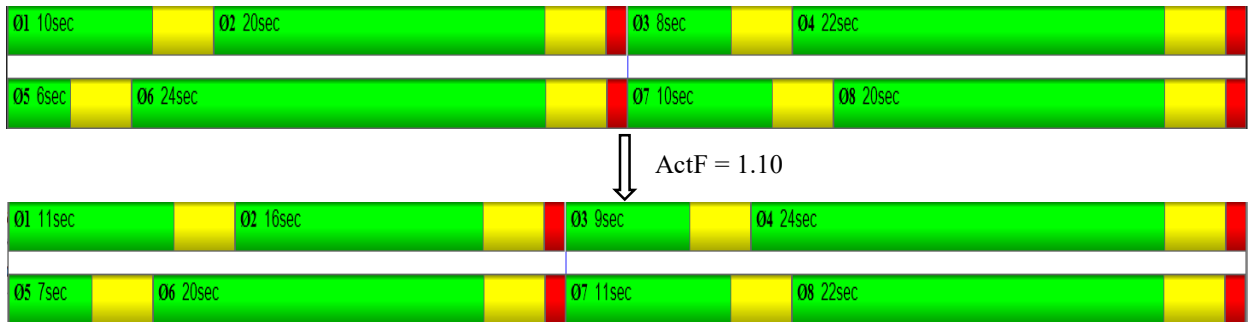

**FIGURE 1** Obtaining the Factored Timing Plan from the Base Timing Plan generated by VISTRO using the actuated factor equal to 1.10.

### Statistical Analysis
The specific objective of this study is to investigate whether applying the ActF to the signal timing plan offered by VISTRO can improve the arterial performance measure. Furthermore, the study seeks for determining an optimal value for the ActF under different levels of traffic demand. If the hypothesis of the existence of such a factor holds true, then the practitioners can implement the optimal value of ActF to the timing plans offered by VISTRO (or any other signal timing optimization tool) and expect a better arterial performance. To achieve this goal, a statistical model requires to be developed to capture the variation of a defined performance index according to distinct values of ActF and various traffic volumes in the arterial. Such a goal can be accomplished through regression analysis. The response variable in the regression model is denoted by Delay Reduction Percentage (DRP) and is formulated in equation (4).

$$DRP = \frac{\text{Average Delay based on Base TP} - \text{Average Delay based on Factored TP}}{\text{Average Delay based on Base TP}} \times 100 \quad (4)$$

The above equation represents the percentage in the delay reduction when the arterial is operated under the Factored TP obtained by the proposed method, in comparison to the Base TP calculated by VISTRO. To account for the traffic fluctuations, the proposed regression model comprises the external traffic volumes as the explanatory variables, in addition to the left-turn rates and ActF.

**Study Site**
An arterial with three intersections located in Prices Fork Road, Blacksburg, Virginia is selected as the test site, as illustrated in Figure 2. All intersections are assumed to be operating by 8 phases with protected left-turn signal configuration on all approaches. Eastbound and westbound directions are the coordinated routes. According to the NEMA phasing configuration, phases 2 and 6 are the coordinated phases along the major street. Traffic volumes of external movements on the side and the major street are a group of explanatory variables in the statistical model that are presented in Figure 2. To compute the traffic volume on each movement using the external volumes, the rate of both the left-turn and right-turn volumes must be identified. In this study, a fixed value equal to 10% of the total volume of an approach is considered as the right-turn proportion while the left-turn proportion is involved as another explanatory variable of the regression model. In summary, the actuated factor (ActF), left-turn rates ($LT$), and external traffic volumes are all selected to be the explanatory variables while DRP in equation (4) is the response variable.

**Actuated Settings**
Since the speed limit of the side streets and left-turn movements is low (less than 35 mph), the detection zones are assumed to be located only at the stop lines. The actuated timing plan settings were selected according to the values recommended by Traffic Signal Timing Manual (*1*). Assuming the speed limits on the side streets and the major street equal to 25 and 35 mph respectively, the minimum green time for the coordinated phases and non-coordinated phases are set as 2 and 7 seconds. This arrangement meets the drivers' expectancy *(1)*.

Yellow change and red clearance intervals are set to 3 and 1 seconds. Following the guidelines of the Traffic Signal Timing Manual and careful observation of simulation runs lead to choosing 2 seconds as the most suitable value for the vehicle extension parameter.

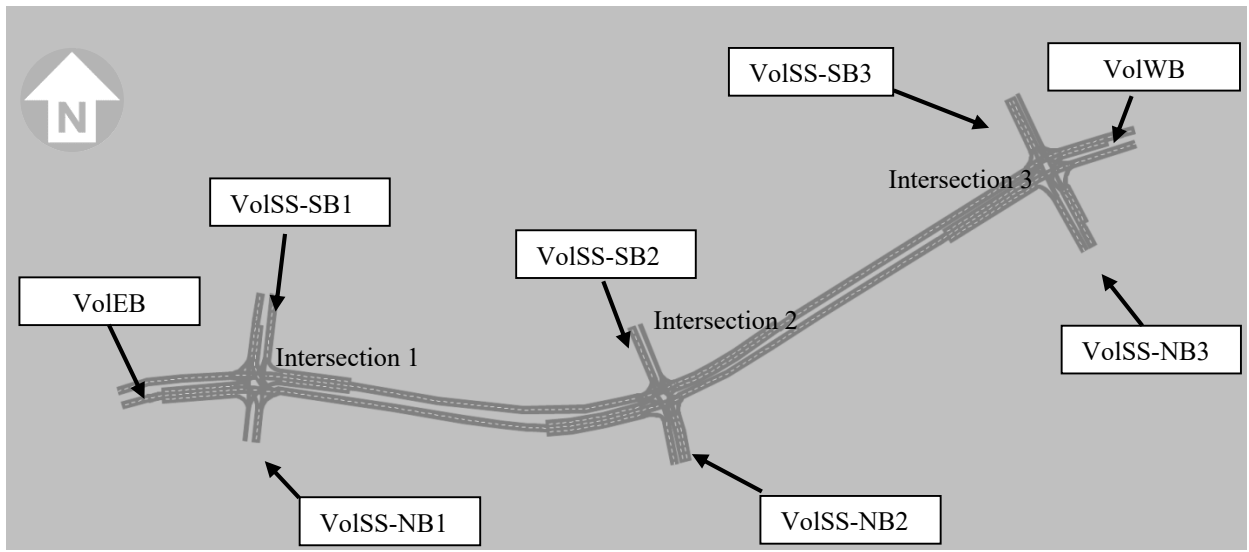

**FIGURE 2  Real-world arterial with three intersections, and the external traffic volumes (explanatory variables) in the arterial.**

**Data Description**
For examining the impact of ActF on the arterial performance measure (e.g., DRP) for various traffic conditions, we need to implement a regression analysis. It is worth mentioning that the purpose of developing a regression analysis is not to predict the DPR, yet to purely test the effect of ActF on DPR. However, we need to first collect data for performing every type of regression models. The first step in the process of data collection for our problem is to perform Design of Experiments (DOE) using JMP Pro 11.0.0 statistical software package. DOE is a cost-effective statistical method which is contingent on a response variable, explanatory variables and the type of the model *(25)*. In this study, the Response Surface Model (RSM) is selected for conducting the experiment. In addition to the explanatory variables (i.e., ActF, *LT*, and external traffic volumes in Table 1), RSM contains two-factor interactions (multiplication of every two variables) and quadratic terms of all explanatory variables.

According to the type and number of factors as well as the chosen model, DOE suggests a specific number of runs. For each experimental run in DOE, the response variable (i.e., DPR) must be computed according to specific values of the explanatory variables. The external volumes for the side streets are continuous variables ranging from 200 to 600 vehicles/hour while the westbound and eastbound traffic volumes in the major street range continuously from 400 to 1200 vehicles/hour. Left-turn percentage (LT) is a discrete variable including three levels: 10%, 20%, and 30%. ActF is also a discrete variable with 7 levels from 1 to 1.30 in 0.05 increments. To minimize the possibility of the green splits of coordinated phases in the Factored TP being lower than their minimum values, the values of ActF are constrained to 1.30 as the upper bound. The number of scenarios recommended by JMP Pro for such a DOE is 72. Each scenario, (i.e., each combination of the variables), is a run with 10 different random seeds to account for the network stochasticity. Therefore, considering the base scenarios operated under the VISTRO signal timing plans, a total of 1440 runs are simulated. The description of the basic explanatory variables used in this study is presented in Table 1. The simulation runs are performed in VISSIM, a microscopic traffic simulation model, to obtain the average delay (seconds/vehicle) as the arterial performance index. After modeling the study site in VISTRO, the following steps are taken in order to compute the response variable for each experimental simulation run:

1. Having the external volumes and LT rate, calculate the traffic volumes for all phases at each intersection.
2. By importing the calculated volumes to VISTRO, generate the base optimal timing plan parameters including cycle length, offsets, and green splits.
3. Determine the Factored TP in accordance with equations (1) to (3).
4. Employing the Ring Barrier Controller, first, import the calculated volumes and the created test site from VISTRO to VISSIM; then, run the simulation for the Base TP and Factored TP in 1-hour analysis period. Note that the simulation network is warmed up for 15 minutes before this 1-hour analysis duration.
5. Record the average delay (seconds/vehicle) corresponding to the Base TP and Factored TP.
6. By using the average delays in step 4, calculate DRP using the equation (4).

**TABLE 1 Description of the Explanatory Variables**

| Variable | Description | Type | Input values | Unit |
| --- | --- | --- | --- | --- |
| VolSS_SBi | External traffic volume, southbound toward intersection i (i = 1, 2, 3). | continuous | 200 to 600 | vehicles/hour |
| VolSS_NBi | External traffic volume, northbound toward intersection i (i = 1, 2, 3). | continuous | 200 to 600 | vehicles/hour |
| VolEB | External traffic volume entering the simulation network from the west side of the arterial. | continuous | 400 to 1200 | vehicles/hour |
| VolWB | External traffic volume entering the simulation network from the east side of the arterial. | continuous | 400 to 1200 | vehicles/hour |
| LT | The left turn rate | discrete | 10, 20, 30 | percent (unitless) |
| ActF | Actuated Factor: Used for defining the amount of the green time to be transferred to non-coordinated phases from the coordinate phases of the signal timing plan optimized by VISTRO. | discrete | 1.00 to 1.30 | - |

**Model Calibration**
After modeling the arterial in VISSIM and before simulating all the DOE scenarios, volume calibration was undertaken for one scenario, called the parent scenario. We build all other scenarios based on the calibrated parent scenario. Travel time counters are coded on all links of the study site to calculate hourly volumes and compare the volumes with the input traffic counts. Volume calibration is achieved by using the GEH formula, invented by Geoffrey E. Havers in the 1970s:

$$GEH = \sqrt{\frac{(E-V)^2}{(E+V)/2}} \qquad (5)$$

where E and V are model estimated volume and traffic counts obtained from DOE, respectively.

The GEH criterion requires the GEH value to be less than 5 and 4 for 85% of the network links and for the entire network, respectively (26). Table below summarizes the results of the volume calibration of the parent scenario, which indicates that the parent scenario has satisfied all requirements.

**TABLE 2  Calibration Results of the Parent Scenario; NA = Not Applicable**

| Link Description | Count | Output | GEH | GEH < 5 | GEH<4 |
|---|---|---|---|---|---|
| Intersection 1 - Eastbound | 826 | 832 | 0.10 | Yes | NA |
| Intersection 1 - Northbound | 215 | 220 | 0.17 | Yes | NA |
| Intersection 1 - Southbound | 636 | 635 | 0.02 | Yes | NA |
| Intersection 1 - Westbound | 333 | 331 | 0.05 | Yes | NA |
| Intersection 3 - Eastbound | 448 | 454 | 0.14 | Yes | NA |
| Intersection 2 - Northbound | 217 | 212 | 0.17 | Yes | NA |
| Intersection 2 - Southbound | 652 | 660 | 0.16 | Yes | NA |
| Intersection 3 - Northbound | 217 | 221 | 0.14 | Yes | NA |
| Intersection 3 - Westbound | 434 | 426 | 0.19 | Yes | NA |
| Intersection 3 - Southbound | 217 | 212 | 0.17 | Yes | NA |
| Between intersections 1 & 2 - Eastbound | 678 | 706 | 0.53 | Yes | NA |
| Between intersections 1 & 2 - Westbound | 339 | 334 | 0.14 | Yes | NA |
| Between intersections 2 & 3 - Eastbound | 609 | 625 | 0.32 | Yes | NA |
| Between intersections 2 & 3 - Westbound | 348 | 343 | 0.13 | Yes | NA |
| Sum | 6169 | 6211 | 0.27 | NA | Yes |

**RESULTS AND DISCUSSION**
Once the experiment procedure was completed by collecting the DPR for all the 72 simulation runs, we develop a regression analysis using JMP Pro 11.0.0. As DOE is based on RSM, the main effects, the two-way interactions, and the quadratic terms were needed to be added in the regression model.

Table 3 reports the *F*-test results of the regression analysis. For the sake of brevity, only the main effects of explanatory variables and ActF$^2$ are represented in Table 3 since all the other variables yield insignificant values. *F*-test examines the hypothesis that all the parameters associated with one particular effect in the model are zero; otherwise, the parameter would have the significant effect in the model. Results of the *F*-test demonstrate that the only significant factor affecting the percentage of delay reduction is the quadratic term of ActF. A very small *p*-value indicates that the term (ActF – 1.15)$^2$ is highly significant even for the significance level of 0.01 as opposed to all combinations of the traffic volumes and the left-turn rate variables that are insignificant.

**TABLE 3 Results of the *F*-test for the Main Effects and the Significant Variable**

| Source | Number of parameters | Degree of Freedom | Sum of Squares | F Ratio | Prob. > F |
|---|---|---|---|---|---|
| VolEB(400,1200) | 1 | 1 | 23.242 | 0.815 | 0.401 |
| VolWB(400,1200) | 1 | 1 | 1.129 | 0.040 | 0.849 |
| VolSS_NB1(200,600) | 1 | 1 | 0.725 | 0.025 | 0.879 |
| VolSS_SB1(200,600) | 1 | 1 | 0.078 | 0.003 | 0.960 |
| VolSS_NB2(200,600) | 1 | 1 | 25.888 | 0.908 | 0.377 |
| VolSS_SB2(200,600) | 1 | 1 | 7.348 | 0.258 | 0.630 |
| VolSS_NB3(200,600) | 1 | 1 | 26.546 | 0.931 | 0.372 |
| VolSS_SB3(200,600) | 1 | 1 | 2.091 | 0.073 | 0.796 |
| LT(10,30) | 1 | 1 | 22.707 | 0.796 | 0.407 |
| ActF (1,1.3) | 1 | 1 | 33.011 | 1.158 | 0.323 |
| **ActF$^2$** | 1 | 1 | 471.447 | **16.535** | **0.007** |

Note: Bold numbers indicate significant variables in the 95% confidence interval.

To better realize and visualize the model and the interesting findings regarding the existence of an optimal value for the actuated factor, Figure 3 illustrates the prediction profilers for all explanatory variables. The prediction profile of the explanatory variable X depicts how the response variable varies when the X variable changes, as shown with black lines in Figure 3. The vertical and horizontal red-dotted lines depict the current values for the explanatory variable X and the response variable. Figure 3 includes five different combinations of traffic volumes on the major and side streets as examples of various traffic conditions. Figures 3a and 3e show the minimum and maximum traffic volumes of all external movements, respectively. Figure 3b presents the maximum traffic volumes for the major street and the minimum traffic volumes for the side streets while the traffic combination of Figure 3c is the other way around of Figure 3b. Finally, Figure 3d is a representative of medium-level traffic volumes in all approaches. The region of the optimal ActF has been highlighted in the very right block of Figures 3a to 3e. Notwithstanding that the traffic fluctuations considerably affect the basic timing plans (i.e., the Base TP), they have a negligible impact on the optimal value of ActF. As can be seen in Figures 3a to 3e, the optimal factor is either 1.15 or very close to 1.15 for our test site. Comparing DPR with ActF = 1.15 and the exact optimal ActF indicates the delay reduction for the ActF of 1.15 and the exact optimal ActF is nearly the same. Therefore, it can be concluded that the optimal ActF value of 1.15 is appropriate for a wide range of traffic conditions. Furthermore, the value of 1.15 is not a large value to decrease the green split of the coordinated phases beyond their minimum green time.

Another interesting finding is that the proposed method is more efficient for the arterials with medium traffic volumes, (i.e., neither operating under the congested nor free flow conditions). Accordingly, the highest values of delay reduction (red values highlighted with green ovals) are corresponding to traffic patterns in Figures 3b, 3c, and 3d. Nonetheless, with regards to Figures 3a and 3e, the proposed method is less efficient when both of the side and the major streets are subjected to either low or high levels of traffic volume.

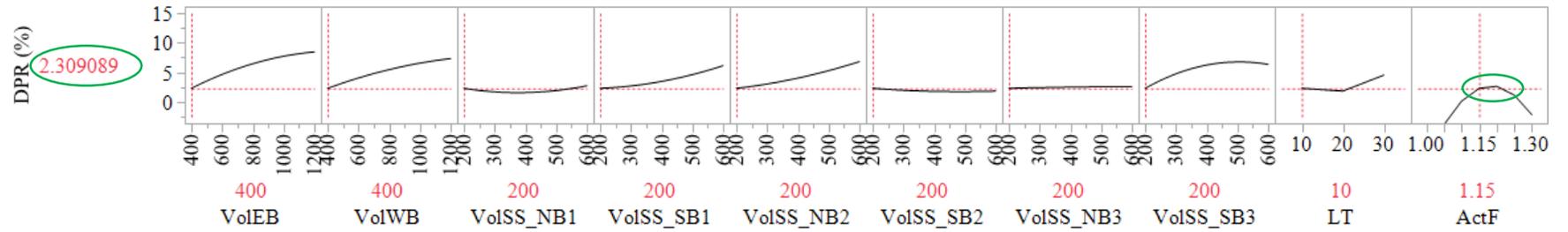

(a)

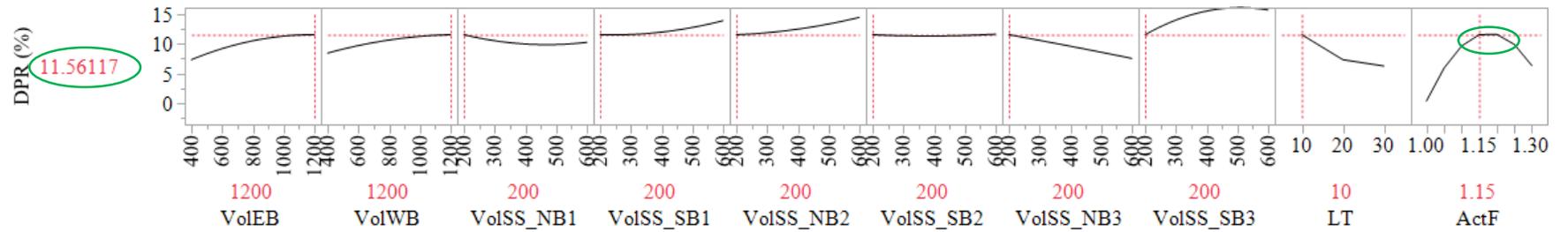

(b)

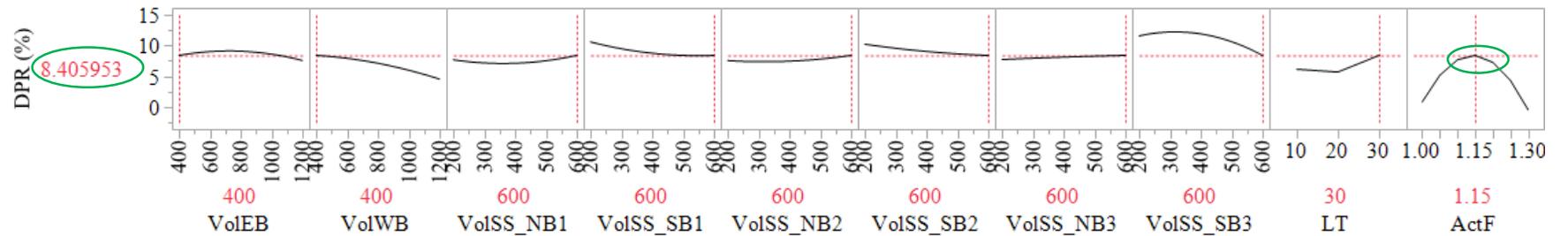

(c)

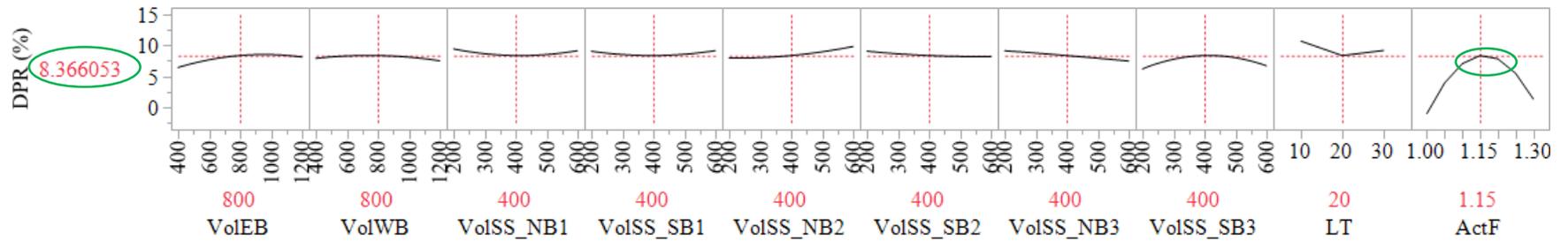

3 (d)

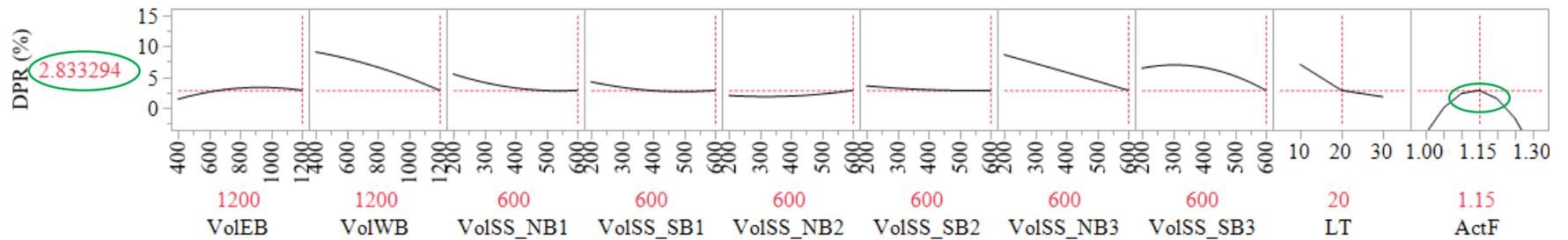

5 (e)

6 **FIGURE 3 Prediction profilers for all the explanatory variables: (a) minimum volumes in all approaches; (b) maximum**
7 **volumes on the major street and minimum volumes on the side streets; (c) minimum volumes on the major street and maximum**
8 **volumes on the side street; (d) moderate volumes in all approaches; (e) maximum volumes in all approaches.**

Dabiri, Kompany, Abbas    13

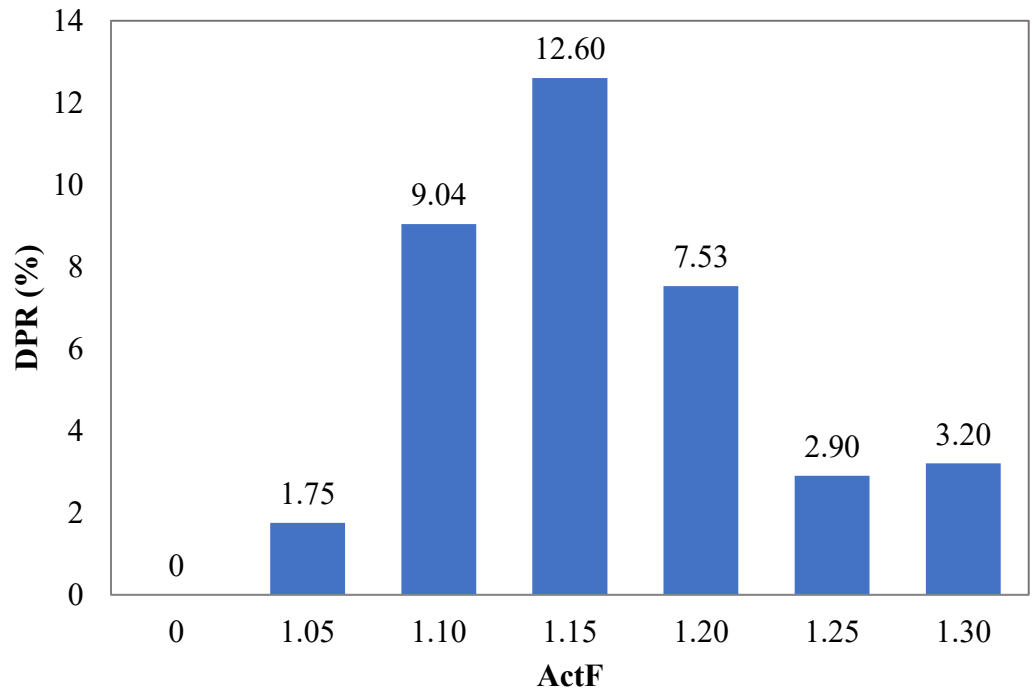

**FIGURE 4  Average DPR(%) versus Actuated Factor (ActF).**

**CONCLUSION**
In this paper, we proposed an efficient and cost-effective procedure to improve the traffic arterial performance operating under the semi-actuated coordinated signal control method. Our method does not require neither installation of detectors for the coordinated movements nor dynamic optimization systems. Furthermore, by having the optimal value of ActF in hand, implementing our method is simple enough to be adopted by practitioners.

The proposed method only changes optimal green splits while keeping the cycle length and offsets unchanged. The non-coordinated phases are multiplied by ActF, which is a factor greater than one. In the meantime, the extra added amount of time in the non-coordinated phases is deducted from the coordinated phases to keep the cycle length constant. Thus, if the traffic demand on the side streets exceeds the threshold for what the basic timing plan (Base TP) was designed, the added time in the non-coordinated phase, obtained through the proposed method, enables the non-coordinated phases to accommodate the additional traffic demand. In pursuit of the proposed procedure, the specific purposes of this study were to not only investigate the efficacy of the ActF for reducing the arterial delay but also determine the optimal ActF for various traffic volumes ranging from uncongested to congested conditions.

To attain the aim of the study and identify in what situations the proposed method results in a higher amount of average delay reduction (seconds/vehicle), a regression analysis was conducted using JMP Pro 11.0.0. In the regression model, the response variable was the amount of delay reduction while the explanatory variables were composed of ActF, external traffic volumes on the side and the major street, as well as the left-turn rate. The statistical analysis indicated that the ActF of 1.15 is a reasonable value for all traffic conditions. Furthermore, the prediction profilers revealed that the proposed method is more effective in reducing delay when the arterial is not operating under the congested or free flow traffic conditions. It is worth mentioning that the regression analysis was conducted to only examine the capability of using the ActF for improving the arterial performance, and not for obtaining a delay formula.

The scope of this study was limited in terms of an arterial with the specific geometry conditions, (e.g., the number of lanes and the number of intersections), and the pre-defined traffic characteristics, (e.g., traffic mode, speed distribution, and PHF). Moreover, in terms of traffic flow conditions, our reported results are applicable as long as traffic volumes in major and minor approaches do not exceed the maximum traffic volumes in Table 1. Future studies need to be carried out to investigate the robustness of the proposed method under a variety of traffic conditions and multi-modal traffic situations. Since considering pedestrians adds more complexity to the semi-actuated coordinated signal control, accounting for pedestrian calls while using our proposed method would be an interesting research area to explore in the future. Moreover, developing a universal regression model that can report the optimal value of the ActF in all traffic and network conditions would be of utmost interest.